\documentclass[a4paper,11pt]{article}
\pdfoutput=1 
\usepackage{amssymb}
\usepackage{amsmath}
\usepackage{braket}
\usepackage{mathtools}
\usepackage[usenames,dvipsnames,svgnames,table]{xcolor}
\usepackage [utf8] {inputenc}
\usepackage{color}
\usepackage{subcaption}
\usepackage{lmodern}
\usepackage{footnote}
\usepackage{slashed}
\usepackage[pdftex] {graphicx}
\usepackage{multirow}
\usepackage{jheppub} 
\usepackage{here}
\usepackage{hyperref}
\usepackage[justification=centering, singlelinecheck=false]{caption}
\usepackage[list=true, labelfont=bf, labelformat=brace, position=top]{subcaption}

\newcommand{\eq}{\begin{eqnarray}}

\newcommand{\en}{\end{eqnarray}}

\title{On the three-particle analog of the Lellouch-L\"uscher formula}
\author[1]{Fabian M\"uller}
\affiliation[1]{Helmholtz-Institut f\"ur Strahlen- und Kernphysik (Theorie) and Bethe Center for Theoretical\\ Physics, Universit\"at Bonn, 53115 Bonn, Germany}

\emailAdd{f.mueller@hiskp.uni-bonn.de}

\author[1,2]{and Akaki Rusetsky}
\affiliation[2]{Tbilisi State  University,  0186 Tbilisi, Georgia}

\emailAdd{rusetsky@hiskp.uni-bonn.de}

\abstract{
  Using  non-relativistic effective field theory, we derive a three-particle analog
  of the Lellouch-L\"uscher formula at the leading order. This formula
  relates the three-particle decay amplitudes in a finite volume with their infinite-volume counterparts and,
  hence, can be used to study the three-particle decays on the lattice. The generalization
of the approach to higher orders is briefly discussed.
}

\allowdisplaybreaks
\begin{document}
\maketitle
\flushbottom

\section{Introduction}

Back in 2000, Lellouch and L\"uscher~\cite{Lellouch:2000pv} derived a formula, which
related the matrix element of the weak $K\to 2\pi$ decay in a finite volume to its
infinite-volume counterpart. These two quantities turn out to be proportional with a factor
(Lellouch-L\"uscher (LL) factor), depending on the size $L$ of a cubic box and on the elastic
two-body pion-pion scattering phase shift.
The result of Ref.~\cite{Lellouch:2000pv}
paved the way to the systematic studies
of various two-body decays on the lattice. Later, different generalizations of the
method emerged, e.g., for
moving frames~\cite{Kim:2005gf,Christ:2005gi}, or for the case of
coupled two-body channels~\cite{Hansen:2012tf}. A simple and transparent
derivation of the LL formula with the use of the non-relativistic effective
Lagrangians has been given in~\cite{Bernard:2012bi}.
For the application of the formalism, we refer here, e.g.,  to a
comprehensive study of the $K\to\pi\pi$ decays, which has been carried
out recently by the RBC and UKQCD Collaborations~\cite{Abbott:2020hxn}.
From the related work, we mention
the study of the matrix elements of currents, corresponding to the $1\to 2$
transition~\cite{Briceno:2015csa,Briceno:2014uqa}, and 
of the timelike pion form factor~\cite{Meyer:2011um},
which all feature the similar factor in a finite volume.
Generally, in the LL type formulae,
this $L$-dependent factor emerges from the multiple rescattering of two particles in the
final state (pions), and the phase shift, which also enters the expression, should be
measured on the same lattice, simultaneously with the measurement of the decay matrix
element. It can be done by using the L\"uscher formula that relates the phase
shift to the volume-dependent spectrum in the two-particle sector.

To summarize, the two-body problem is completely understood from the conceptual point
of view -- both the scattering, as well as two-body decays. On the contrary, the three-body
formalism is still in development. Recently, three physically equivalent forms of the quantization condition
have been proposed~\cite{Hansen:2014eka,Hansen:2015zga,Hammer:2017uqm,Hammer:2017kms,Mai:2017bge}, which relate the three-body spectrum in a finite
volume with the infinite-volume parameters in the three-body sector.
However, in contrast to the
two-body case, where the L\"uscher equation enables one to extract the two-body phase
shift from the measured spectrum in one step, the procedure in the three-particle
sector is more complicated. To start with, the three-body quantization condition
becomes tractable only if the three-body interactions are expressed in terms of few
parameters. In the approach of Refs.~\cite{Hammer:2017uqm,Hammer:2017kms}, such
a parameterization naturally emerges, when the three-body interactions are evaluated from
the effective Lagrangian at tree level that allows one to impose a consistent power
counting. Similarly, the three-body kernels in the approaches of
Refs.~\cite{Hansen:2014eka,Hansen:2015zga} and
Ref.~\cite{Mai:2017bge} can be expanded
in the external momenta (up to a given order) in the vicinity of the threshold.
Thus, the fit of the quantization condition to the 
three-particle spectrum, which is measured on the lattice,  enables one to extract
few parameters (the effective three-particle couplings, or the coefficients in the
expansion of the three-particle kernel). These can be substituted into
the
infinite-volume equations to calculate observables in the three-particle sector.
Consequently,
extracting the three-particle observables from data necessarily involves an intermediate
step, and cannot be done directly, as in case of two particles.

It should be mentioned that
the above theoretical developments have largely boosted the study
of three (and more particles) on the lattice, be this in QCD or other
field-theoretical models~\cite{Beane:2007es,Horz:2019rrn,Culver:2019vvu,Fischer:2020jzp,Blanton:2019vdk,Hansen:2020otl,Alexandru:2020xqf,Romero-Lopez:2018rcb,Romero-Lopez:2020rdq}.
In view of these activities, the need for the three-particle analog
of the LL formula, which should be used for the extraction of the matrix
elements, becomes obvious. Such a formula, however, was not available
in the literature so far. Moreover,
bearing in mind the above discussion, it is not even clear, whether the
relation between the finite- and infinite-volume matrix elements, which one is after,
should contain a single overall factor (a counterpart of the LL factor),
or should be more complicated. On the other hand, recent years have seen a growing
interest to the study of three-particle decays. The most obvious candidates for this study
in the beginning
are provided by the three-pion decays of low-mass light-flavored mesons $K\to 3\pi$,
$\eta\to 3\pi$ and $\omega\to 3\pi$. The decays of the heavier pseudovector mesons
$a_1(1260)\to\rho\pi\to 3\pi$ and $a_1(1420)\to f_0(980)\pi\to 3\pi$ are also very
interesting\footnote{As one will see later, in the lattice study of
all these decays,
a prior knowledge of the three-pion amplitude is necessary. The total isospin
of the decay products in the above processes is different, but neither of
them equals the maximal possible isospin $I=3$, available in the system of
three pions. It is important to note that
the three-body finite-volume formalism, which enables one to explore the
systems with an arbitrary isospin, has become available only
very recently~\cite{Hansen:2020zhy,Muller:2020vtt}.}. Further,
the candidates for exotica, $X(3872)$ and $X(4260)$, decay largely into the
three-particle final states as well. Last but not least, the extraction of the parameters of
the Roper resonance on the lattice has proven to be very challenging.
That might be, in part,
related to the lack of proper treatment of the three-particle decay channel in a
finite volume. Our paper intends to make the first step towards the creation of a systematic
finite-volume framework for the study of three-body decays on the lattice that
will contribute to the solution of the above-mentioned problems\footnote{After the present paper was submitted to the archive, Ref.~\cite{Hansen:2021ofl}, which deals with the same issue, has appeared.}.

Note that the resonances, which are studied in lattice QCD, fall into two
categories. To the first category belong the ones, which are stable in pure QCD, like kaons that decay through weak interactions.
Further, the $\eta$-mesons are not stable in QCD. However, the decay
amplitude is proportional to the $u$- and $d$-quark mass difference $m_u-m_d$
and thus vanishes in the isospin limit. So, if one wants to know this amplitude
only at the first order in $m_u-m_d$ (this completely suffices for practical
reasons), one could also formally
categorize this decay into the first group and
treat the final state interactions in the isospin-symmetric QCD, where the
$\eta$-mesons are stable. The second, larger group
consists of the genuine QCD resonances. In this paper, like
in the original paper by
Lellouch and L\"uscher, we concentrate our effort on the first group.
The treatment of the QCD resonances is a more subtle
issue that includes, in particular, analytic continuation into the complex energy plane
to the resonance pole. In the case of two-body decays, this procedure is discussed,
in particular, in Refs.~\cite{Bernard:2012bi,Agadjanov:2014kha,Agadjanov:2016fbd,Briceno:2015dca,Briceno:2016kkp}.
We postpone the discussion of a similar procedure in the three-particle sector
to our future investigations.

The layout of the paper is as follows: In Sect.~\ref{sec:Lagrangian}, we display the
lowest-order non-relativistic effective Lagrangian and write down the quantization
condition in a finite volume. In Sect.~\ref{sec:LL}, we derive the LL
equation at leading order. The extension of the approach to higher orders is discussed
in Sect.~\ref{sec:high}. Section~\ref{sec:concl} contains our conclusions.

\section{Non-relativistic framework}
\label{sec:Lagrangian}

The non-relativistic EFT framework, which was tailored to study the singularity structure
of the amplitudes in three-body decays, was proposed in
Ref.~\cite{Colangelo:2006va}. It has been successfully used in the study of
the three-body decays of charged and neutral kaons, as well as $\omega,\eta$ and $\eta'$
mesons~\cite{Bissegger:2008ff,Bissegger:2007yq,Gullstrom:2008sy,Schneider:2010hs,Kubis:2009sb}. A brief review of the essential points of the approach is given
in Ref.~\cite{Gasser:2011ju}. The main difference of this approach from the conventional
ones consists in the treatment of the relativistic corrections to the internal particle lines.
Whereas in the conventional approach, these corrections are treated perturbatively,
in the new one they are summed up to all orders, ensuring the correct relativistic
dispersion law. As a result, the location of singularities in the decay amplitude
stays fixed to all orders and coincides with the singularity structure of the relativistic
amplitude. There is a price to pay for this, however: the resummed propagators feature
the hard scale -- the particle mass -- explicitly. This, as known, leads to the breakdown of
the naive counting rules. In order to rectify the counting rules, one then has to amend
the procedure for the calculation of the Feynman integrals -- dimensional
regularization plus minimal subtraction does not suffice. The modification of the procedure,
which is equivalent to the change of the renormalization prescription, is described
in detail in~\cite{Gasser:2011ju}, and we refer an interested reader to that article.

To purify the problem from the inessential details as much as possible, we shall consider
below a decay of a spinless particle (``kaon'') into three likewise spinless particles
(``pions''). Isospin and other quantum numbers are discarded. We also assume that there
exists some discrete symmetry (like $G$-parity), which forbids transitions with an odd
number of the external pion legs.
The non-relativistic fields $K(x)$ and $\phi(x)$ describe kaons and pions, respectively,
and $M,m$ denote their masses. The lowest-order Lagrangian, which describes the decay,
is given by
\eq\label{eq:Lagr-LO}
\mathcal{L} &=&
K^\dagger 2W(i\partial_t-W)K
+\phi^\dagger 2w(i\partial_t - w) \phi
\nonumber\\[2mm]
&+& \frac{C_0}{4} \phi^\dagger \phi^\dagger \phi \phi
+ \frac{D_0}{36} \phi^\dagger \phi^\dagger \phi^\dagger \phi \phi \phi
+ \frac{G_0}{6} \left( K^\dagger \phi \phi \phi + \text{h.c.} \right)\,,
\en
where $W=\sqrt{M^2-\nabla^2}$ and $w=\sqrt{m^2-\nabla^2}$.

In the above Lagrangian, the constant $G_0$ describes the elementary act of the kaon
decay into three pions. It is proportional to the weak coupling constant and enters
 the amplitudes, by definition, only at the first order. If the weak interactions are switched
off, the kaon is stable. Further, the constants $C_0$ and $D_0$ describe the strong
final state
interactions in the system of two and three pions respectively. Unlike the constant $G_0$,
these enter the expression of the amplitude at all orders. The matching at the two-pion
threshold relates the constant $C_0$ to the pion-pion scattering length $a$:
\eq
C_0=-32\pi am\, .
\en
In Eq.~(\ref{eq:Lagr-LO}), only the leading-order terms are displayed. The power counting
at the Lagrangian level is defined by the (formal)
requirement that all three-momenta count
at $O(p)$, whereas the kinetic energies of the individual pions, as well as the quantity
$M-3m$ count at $O(p^2)$. The higher-order Lagrangians would contain an even number of
spatial derivatives, acting on  all fields. Below, we shall concentrate on the derivation
of the LL formula at the leading order, using the Lagrangian in Eq.~(\ref{eq:Lagr-LO}).
The inclusion of the higher-order terms will be considered briefly in Sect.~\ref{sec:high}.

Moreover, in order to write down the equation that determines the three-particle
scattering amplitude, we shall switch to the particle-dimer picture. It is
well-known that
this formulation, which is equivalent to the original one, enables one to drastically simplify
the bookkeeping of Feynman diagrams and arrive at the result with a surprising
ease~\cite{Kaplan:1996nv,Bedaque:1998kg,Bedaque:1998km}. The Lagrangian in the
particle-dimer picture in our case is given by:
\eq\label{eq:L-dim}
\mathcal{L} &=&
K^\dagger 2W(i\partial_t-W)K
+\phi^\dagger 2w(i\partial_t - w) \phi + \sigma d^\dagger d
\nonumber\\[2mm]
&+& \frac{f_0}{2} \left( d^\dagger \phi \phi + \text{h.c.} \right)
+ h_0 d^\dagger d \phi^\dagger \phi + g_0 \left(K^\dagger d \phi + \text{h.c.} \right)\, .
\en
Here, $d$ denotes the dimer field, and $\sigma=\pm 1$, depending on the sign of the constant $C_0$. Integrating out the field $d$ in the path integral and expanding in the powers
of fields, one arrives at the Lagrangian, given in Eq.~(\ref{eq:Lagr-LO}), if the following relations are fulfilled:
\eq
\sigma f_0^2 = - C_0, \qquad 9 f_0^2 h_0 = D_0, \qquad 3 \sigma f_0 g_0 = - G_0\, .
\en
We would like to stress here that the validity of the particle-dimer picture
does not imply that a two-body bound state really exists. As one sees, the
dimer field is introduced in the path integral as a dummy integration variable
and, hence, the resulting formulation is mathematically equivalent to the initial
one without a dimer field. If a dimer (or a narrow low-lying resonance)
indeed exists, this may affect only the convergence of the expansion. In this
case, the bulk of the two-particle interaction will be described by the
dimer exchange in the $s$-channel, and the contribution from the higher
orders will be small.

\begin{figure}[t]
  \begin{center}
    \includegraphics[width=12cm]{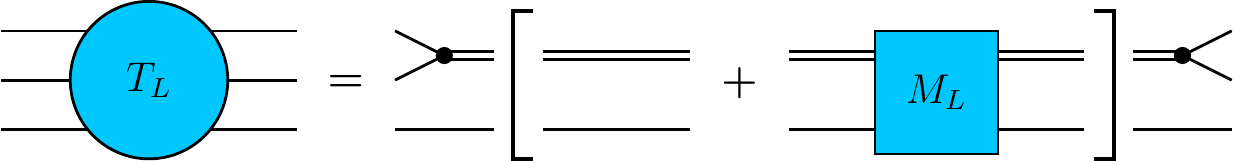}
  \end{center}
    \caption{Expressing the three-particle scattering amplitude through the particle-dimer scattering amplitude, see Eq.~(\ref{eqn:three_particle_scattering}). The single and double lines are the particle and the dimer propagators, respectively, and the filled circle denotes the two-particle-dimer vertex.}
    \label{fig:3-M}
  \end{figure}

The Lagrangian~(\ref{eq:L-dim}) will be used for the calculation of the Feynman diagrams
in a finite volume -- as it is well-known,
the sole change is the replacement of the infinite-momentum integrals
by the sums over the discrete three-momenta of particles in a finite cubic
box\footnote{For simplicity, below we display all formulae in the Minkowski space.
  The final results, obtained with the use of Wick rotation, are identical.}.
The propagator of the non-relativistic field $\phi(x)$ in a finite volume is given by:
\eq
 i\langle 0 | T \phi(x)\phi^\dagger(y) | 0 \rangle =
 \int\frac{dp_0}{2\pi}\phantom{.} \frac{1}{L^3}\sum_{\mathbf{p}}
 \frac{e^{-ip_0(x_0-y_0) + i \mathbf{p}(\mathbf{x}-\mathbf{y})}}
 {2w(\mathbf{p})( w(\mathbf{p}) - p_0-i\varepsilon )}\,,
 \qquad w(\mathbf{p}) = \sqrt{m^2+\mathbf{p}^2}\, .\quad
\en
The dimer propagator is obtained by summing pion loops to all orders:
\eq
 i\langle 0 | T d(x)d^\dagger(y) | 0 \rangle =
 \int\frac{dP_0}{2\pi}\phantom{.} \frac{1}{L^3}\sum_{\mathbf{P}}
 e^{-iP_0(x_0-y_0) + i \mathbf{P}(\mathbf{x}-\mathbf{y})}\,
 D_L(\mathbf{P};P_0)\, .
 \en
Here $D_L$ obeys the following equation:
\eq
D_L(\mathbf{P};P_0) = -\frac{1}{\sigma}
-\frac{f_0^2}{2\sigma}\,J_L(\mathbf{P};P_0)D_L(\mathbf{P};P_0)\, ,
\en
where $J_L$ denotes a single pion loop:
\eq\label{eq:JL}
J_L(\mathbf{P};P_0) &=&
\frac{1}{L^3} \sum_{\mathbf{k}} \frac{1}{4w(\mathbf{k})w(\mathbf{P}-\mathbf{k})( w(\mathbf{k}) + w(\mathbf{P}-\mathbf{k}) -P_0-i\varepsilon)}
\nonumber\\[2mm]
&=& \frac{p^*}{8\pi^{5/2}\sqrt{s}\gamma\eta}Z^{\mathbf{d}}_{00}(1;s)\, ,
\en
and
\eq
s =P_0^2-\mathbf{P^2},\quad \gamma=\frac{P_0}{\sqrt{s}}\,,\quad
p^*=\sqrt{\frac{s}{4}-m^2}, \quad \eta=\frac{p^* L}{2\pi}\, ,\quad
\mathbf{d}=\frac{\mathbf{P}L}{2\pi}\, .
\en
Further, in Eq.~(\ref{eq:JL}), $Z^{\mathbf{d}}_{00}(1;s)$ is the usual L\"uscher zeta function, boosted to the moving frame defined by the vector $\mathbf{d}$. For a general
$(lm)$, this function is given by:
\eq
Z^{\mathbf{d}}_{lm}(1;s)=\sum_{\mathbf{r}\in P_d}
\frac{\mathcal{Y}_{lm}(\mathbf{r})}{\mathbf{r}^2-\eta^2}\, ,\quad
P_d=\left\{\mathbf{r}\in\mathbb{R}^3\biggl|r_\parallel=\gamma^{-1}\left(n_\parallel-|{\bf d}|/2\right)\, ,~\mathbf{r}_\perp=\mathbf{n}_\perp\, ,~\mathbf{n}\in\mathbb{Z}^3\right\}\, ,
\nonumber\\
\en
where $\mathcal{Y}_{lm}(\mathbf{r})=|\mathbf{r}|^lY_{lm}(\hat r)$, and $Y_{lm}(\hat r)$
denotes the usual spherical function that depends on the unit vector $\hat r$.
Finally, after using the matching condition, for the dimer propagator one obtains:
\eq\label{eq:DL}
D_L(\mathbf{P};P_0) = \frac{\sigma\sqrt{s}/(2am)}
  {-\sqrt{s}/(2am) + p^*\cot\phi^\mathbf{d}(s)}, \quad\quad
  \cot\phi^\mathbf{d}(s) = -\frac{Z_{00}^\mathbf{d}(1;s)}{\pi^{3/2}\gamma\eta}\, .
\en
In the non-relativistic limit, $\sqrt{s}/(2m)\to 1$, $\gamma\to 1$, and we arrive at the expression displayed in Refs.~\cite{Hammer:2017uqm,Hammer:2017kms}.
At higher orders, the expression $-1/a$ both in the numerator and the
denominator gets replaced by $p^*\cot\delta(p^*)=-1/a+r{p^*}^2/2+\ldots$.
Further, the infinite-volume counterpart of Eq.~(\ref{eq:DL}) reads:
\eq\label{eq:D}
D(\mathbf{P};P_0) = \frac{\sigma\sqrt{s}/(2am)}
  {-\sqrt{s}/(2am) -ip^*}\, .
\en

\begin{figure}[t]
  \begin{center}
    \includegraphics[width=12cm]{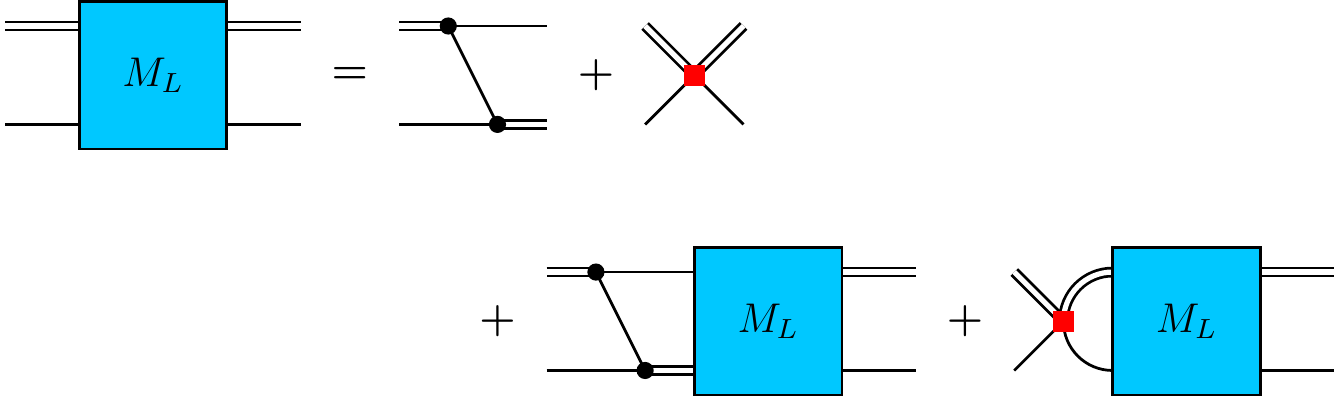}
  \end{center}
    \caption{The Faddeev equation for the particle-dimer scattering amplitude. The red shaded squares denote the particle-dimer coupling.}
    \label{fig:Faddeev}
  \end{figure}

The finite-volume energy levels in the three-particle system coincide with
the location of the poles of the three-particle scattering amplitude. In the
particle-dimer picture, this quantity can be directly related to the
particle-dimer scattering amplitude~\cite{Hammer:2017uqm,Hammer:2017kms,Doring:2018xxx}, see Fig.~\ref{fig:3-M}. At the lowest order,
the relation is given by:
\eq\label{eqn:three_particle_scattering}
T_L(\{\mathbf{p}\},\{\mathbf{q}\};P_0)
&=& \sum\limits_{\alpha,\beta=1}^3
\biggl[ \tau_L(-\mathbf{p}_\alpha;P_0)2w(\mathbf{p}_\alpha)
  L^3\delta_{\mathbf{p}_\alpha\mathbf{q}_\beta}
\nonumber\\[2mm]
  &+&
  \tau_L(-\mathbf{p}_\alpha;P_0) M_L(-\mathbf{p}_\alpha,-\mathbf{q}_\beta;P_0) \tau_L(-\mathbf{q}_\beta;P_0)\biggr]\, ,
\en
where $\{\mathbf{p}\}$ stands for the set of all three
particle momenta $\mathbf{p}_\alpha$ with $\alpha=1,2,3$.
In the center-of-mass frame, the dimer momenta are equal to $-\mathbf{p}_\alpha$.
The sets $\{\mathbf{q}\}$, $\{\mathbf{k}\}$ are defined similarly.
Further,
\eq
\tau_L(\mathbf{p};P_0)=f_0^2D_L(\mathbf{p};P_0-w(\mathbf{p}))\, .
\en
In the infinite volume, the relation between the quantities $\tau(\mathbf{p};P_0)$
and $D(\mathbf{p};P_0)$ takes a similar form. 
Further, $M_L(\mathbf{p},\mathbf{q};P_0)$ denotes the particle-dimer
scattering amplitude, which obeys the Faddeev
equation in a finite volume, see Fig.~\ref{fig:Faddeev}:
\eq
M_L(\mathbf{p},\mathbf{q};P_0) = Z(\mathbf{p},\mathbf{q};P_0)
+ \frac{1}{L^3}\sum_{\mathbf{k}}^{\Lambda} Z(\mathbf{p},\mathbf{k};P_0)
\frac{\tau_L(\mathbf{k};P_0)}{2w(\mathbf{k})} M_L(\mathbf{k},\mathbf{q};P_0)\, ,
\en
where $\Lambda$ denotes an ultraviolet cutoff and
\eq
Z(\mathbf{p},\mathbf{q};P_0) = \left[ \frac{1}{2w(\mathbf{p}+\mathbf{q})( w(\mathbf{p}+\mathbf{q})+w(\mathbf{p})+w(\mathbf{q})-P_0-i\varepsilon)} + \frac{h_0}{f_0^2}\right]\, .
\en
In the infinite volume, the Faddeev equation becomes the integral equation with the same kernel $Z$ and cutoff $\Lambda$:
\eq\label{eq:int-Faddeev}
M(\mathbf{p},\mathbf{q};P_0) = Z(\mathbf{p},\mathbf{q};P_0) +
\int^\Lambda\frac{d^3\mathbf{k}}{(2\pi)^3}\,
Z(\mathbf{p},\mathbf{k};P_0) \frac{\tau(\mathbf{k};P_0)}{2w(\mathbf{k})}
M(\mathbf{k},\mathbf{q};P_0)\, .
\en
The quantization condition in a finite volume takes the form:
\eq
\mbox{det}(A)=0\, ,\qquad
A_{\mathbf{p}\mathbf{q}}=2w(\mathbf{p})\tau_L^{-1}(\mathbf{p};P_0)L^3\delta_{\mathbf{p}\mathbf{q}}-Z(\mathbf{p},\mathbf{q};P_0)\, .
\en
The discrete solutions $P_0=E_n$ of the quantization condition determine
the finite-volume spectrum of the three-particle system. Further, in the
vicinity of a pole $P_0=E_n$, the residue of the particle-dimer amplitude
factorizes:
\eq\label{eq:ML-factor}
M_L(\mathbf{p},\mathbf{q};P_0)\biggr|_{P_0\to E_n} =
\frac{\psi_L^{(n)}(\mathbf{p})\psi_L^{(n)}(\mathbf{q})}{E_n -P_0}
+ \mbox{regular}\, .
\en
The particle-dimer wave function obeys a homogeneous equation: 
\eq\label{eq:psi_L}
\psi_L^{(n)}(\mathbf{p}) = \frac{1}{L^3} \sum_{\mathbf{k}}^{\Lambda} Z(\mathbf{p},\mathbf{k};E_n) \frac{\tau_L(\mathbf{k};E_n)}{2w(\mathbf{k})} \psi_L^{(n)}(\mathbf{k})\, . 
\en
The normalization condition for the finite-volume wave function $\psi^{(n)}_L({\bf p})$
can be derived in a standard manner by using Eqs.~(\ref{eq:int-Faddeev}),
(\ref{eq:ML-factor}) and (\ref{eq:psi_L}). Since both $Z$ and $\tau_L$ are
energy-dependent, $\psi_L({\bf p})$ is not merely normalized to unity. Instead, the normalization condition takes the form:
\eq
&&\frac{1}{L^6}\sum_{{\bf p},{\bf k}}^\Lambda\psi^{(n)}({\bf p})
\frac{\tau_L(\mathbf{p};E_n)}{2w(\mathbf{p})}\,
\frac{dZ({\bf p},{\bf k};E_n)}{dE_n}\,
\frac{\tau_L(\mathbf{k};E_n)}{2w(\mathbf{k})}\,\psi^{(n)}_L({\bf k})
\nonumber\\[2mm]
&&\quad+\frac{1}{L^3}\sum_{\bf p}^\Lambda\psi^{(n)}({\bf p})
\frac{1}{2w(\mathbf{p})}\,
\frac{d\tau_L(\mathbf{p};E_n)}{dE_n}\,\psi^{(n)}_L({\bf p})=1\, .
\en
The three-particle scattering amplitude factorizes as well:
\eq\label{eq:factorize-T}
T_L(\{\mathbf{p}\},\{\mathbf{q}\};P_0)\biggr|_{P_0\to E_n} =\frac{\Psi_L^{(n)}(\{\mathbf{p}\})\Psi_L^{(n)}(\{\mathbf{q}\})}{E_n -P_0}
+ \mbox{regular}\, ,
\en
where
\eq
\Psi_L^{(n)}(\{\mathbf{p}\})=\sum_{\alpha=1}^3\tau_L(-\mathbf{p}_\alpha;E_n)
\psi_L^{(n)}(-\mathbf{p}_\alpha)\, .
\en
Up to the change to the relativistic normalization and the use of the
relativistic kinematics in the dimer propagator, these equations are
equivalent to the ones displayed in
Refs.~\cite{Hammer:2017uqm,Hammer:2017kms,Doring:2018xxx}.
The numerical solution of similar equations in a finite volume has been considered also, e.g., in Refs.~\cite{Kreuzer:2010ti,Kreuzer:2009jp,Kreuzer:2008bi,Kreuzer:2012sr}.

\section{Derivation of the three-particle analog of the LL formula at the leading
  order}
\label{sec:LL}

The derivation of the counterpart of the LL formula in the three-particle sector proceeds
along the path already used in the two-particle case~\cite{Bernard:2012bi}. The
main idea can be formulated in few words. The non-relativistic effective
Lagrangians, used to describe physics in the infinite and in a finite volume, are the same.
At the leading order, the only unknown, which can be extracted from the measured
$K\to 3\pi$ decay matrix element on the lattice, is the
coupling $G_0$ (other couplings, $C_0$ and $D_0$, can be independently
determined by measuring the two- and three-body energy levels). Hence,
the only thing that one
has to do is to calculate the decay matrix elements in the effective theory twice:
in a finite and in the infinite volume. Since at the leading order this matrix element is
merely proportional to $G_0$, in the ratio of the results of the two calculations, which
is the three-particle analog of the LL factor we are looking for, this constant
drops out. Thus, the final answer is expressed solely in terms of known constants
$C_0$ and $D_0$.

The crucial point in this derivation is to concentrate on $G_0$ which, by definition,
is the same in a finite and in the infinite volume, up to the exponentially suppressed
corrections. In these corrections, the hard scale of the effective theory appears in the argument of the
exponent (in our case, this hard scale is given by the pion mass $m$). On the contrary,
the measured matrix element contains a non-trivial, power-law $L$-dependence, which
emerges via the final state interactions. Hence, no regular $L\to\infty$ limit exists for
this matrix element.

After this introductory remark, we proceed with the calculation of the decay matrix
element. Following Ref.~\cite{Bernard:2012bi}, first, one has to calculate the
wave function renormalization constant for the composite operator
$\mathcal{O}(x_0;\{\mathbf{k}\})$, which creates three pions with momenta
$\mathbf{k}_1,\mathbf{k}_2,\mathbf{k}_3$, acting on the vacuum bra-vector
$\langle 0|$:
\eq
\mathcal{O}(x_0;\{\mathbf{k}\})
= \int d^3\mathbf{x}_1d^3\mathbf{x}_2d^3\mathbf{x}_3\phantom{.}
e^{-i\mathbf{k}_1\mathbf{x}_1-i\mathbf{k}_2\mathbf{x}_2-i\mathbf{k}_3\mathbf{x}_3}
\phi(x_0,\mathbf{x}_1)\phi(x_0,\mathbf{x}_2)\phi(x_0,\mathbf{x}_3)\, .
\en
Assume now that $x_0>y_0$. Inserting a complete set of the intermediate states,
for the two-body correlator one gets:
\eq
\langle 0 | \mathcal{O}(x_0;\{\mathbf{k}\}) \mathcal{O}^\dagger(y_0;\{\mathbf{k}\}) |0\rangle
= \sum_{n} \left| \langle 0| \mathcal{O}(0;\{\mathbf{k}\}) |n\rangle  \right|^2
e^{-iE_n(x_0-y_0)}\, .
\en
On the other hand, one can evaluate this correlator in the perturbation theory.
Summing up all diagrams, one obtains:
\eq
  \label{eqn:matrix_element_perturbation_theory}
 &&\langle 0 | \mathcal{O}(x_0;\{\mathbf{k}\})
  \mathcal{O}^\dagger(y_0;\{\mathbf{k}\}) |0\rangle
  = \int\frac{dP_0}{2\pi i}\phantom{.} e^{-iP_0(x_0-y_0)}
  \nonumber\\[2mm]
&\times&\biggl\{
\frac{L^9\left(1+\delta_{\mathbf{k}_1\mathbf{k}_2}
    +\delta_{\mathbf{k}_1\mathbf{k}_3}+\delta_{\mathbf{k}_2\mathbf{k}_3}
    +2\delta_{\mathbf{k}_1\mathbf{k}_2}\delta_{\mathbf{k}_2\mathbf{k}_3}\right)}
{8w(\mathbf{k}_1)w(\mathbf{k}_2)w(\mathbf{k}_3)
  (w(\mathbf{k}_1)+w(\mathbf{k}_2)+w(\mathbf{k}_3)- P_0-i\varepsilon)}
\nonumber\\[2mm]
&+& \frac{L^3 T_L(\{\mathbf{k}\},\{\mathbf{k}\};P_0)}
{\left(8w(\mathbf{k}_1)w(\mathbf{k}_2)w(\mathbf{k}_3)
    (w(\mathbf{k}_1)+w(\mathbf{k}_2)+w(\mathbf{k}_3)-P_0-i\varepsilon)\right)^2}
\biggr\}\, .
\en
Using Eq.~(\ref{eq:factorize-T}) and performing contour integration by means of the
Cauchy theorem, one gets:
\eq
\langle 0 | \mathcal{O}(x_0;\{\mathbf{k}\})\mathcal{O}^\dagger(y_0;\{\mathbf{k}\}) |0\rangle
= \sum_{n} 
\frac{ e^{-iE_n(x_0-y_0)}L^3 \left(\Psi_L^{(n)}(\{\mathbf{k}\})\right)^2}
{\left(8w(\mathbf{k}_1)w(\mathbf{k}_2)w(\mathbf{k}_3)(w(\mathbf{k}_1)+w(\mathbf{k}_2)+w(\mathbf{k}_3)-E_n)\right)^2}\, .
\nonumber\\
\en
From this, we finally obtain:
\eq\label{eqn:overpal_eigenstates}
|\langle 0 | \mathcal{O}(0;\{\mathbf{k}\}) |n\rangle|
= L^{3/2}
\frac{ \left| \Psi_L^{(n)}(\{\mathbf{k}\})\right|}
{\left|8w(\mathbf{k}_1)w(\mathbf{k}_2)w(\mathbf{k}_3)(w(\mathbf{k}_1)+w(\mathbf{k}_2)+w(\mathbf{k}_3)-E_n)\right|}\, .
\en
In the above derivation, it was assumed that the free-particle singularities, emerging from
the energy denominators in Eq.~(\ref{eqn:matrix_element_perturbation_theory}), cancel
in the full expression for the correlator. This statement, which is evident on general
grounds, was verified (in threshold kinematics) in Ref.~\cite{Polejaeva:2012ut}.
We refer an interested reader to that paper for more details.

\begin{figure}[t]
  \begin{center}
    \includegraphics[width=12cm]{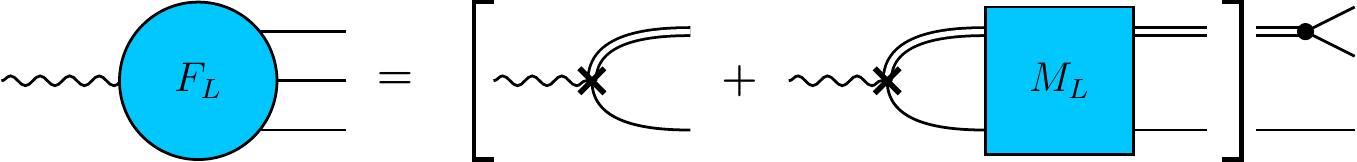}
  \end{center}
  \caption{The amplitude of the $K\to 3\pi$ decay in the particle-dimer picture. The notations are the same as in Figs.~\ref{fig:3-M} and \ref{fig:Faddeev}. The cross denotes the
    vertex, which correspond to the decay of a kaon into a particle-dimer pair. This vertex comes with the coupling $g_0$.}
    \label{fig:vertex}
  \end{figure}

Next, we calculate the decay matrix element. First, note that the kaon interaction term
in the particle-dimer Lagrangian~(\ref{eq:L-dim}) can be rewritten in a form
$J_K^\dagger(x)K(x)+\mbox{h.c.}$, where
\eq
J_K^\dagger=g_0d^\dagger\phi^\dagger\, .
\en
Consequently, on the one hand, assuming $x_0>0$, one gets:
\eq
\label{eqn:three_particle_creation_eigenstates}
\langle 0| \mathcal{O}(x_0;\{ \mathbf{k}\})J^\dagger_K(0) |0\rangle
= \sum_{n} e^{-iE_nx_0}  \langle 0| \mathcal{O}(0;\{ \mathbf{k} \}) |n\rangle \langle n| J^\dagger_K(0)|0\rangle\, .
\en
On the other hand, using perturbation theory and summing up pertinent diagrams
results in:
\eq\label{eq:FL}
\langle 0| \mathcal{O}(x_0;\{ \mathbf{k}\})J^\dagger_K(0) |0\rangle
=\int \frac{dP_0}{2\pi i}\,\frac{e^{-iP_0x_0}F_L(\{\mathbf{k}\};P_0)}{8w(\mathbf{k}_1)w(\mathbf{k}_2)w(\mathbf{k}_3)(w(\mathbf{k}_1)+w(\mathbf{k}_2)+w(\mathbf{k}_3)-P_0-i\varepsilon)}\, ,
\nonumber\\[2mm]
\en
where (see Fig.~\ref{fig:vertex}):
\eq
F_L(\{\mathbf{k}\};P_0)
= \frac{g_0}{f_0} \sum_{\alpha=1}^{3}
\tau_L(-\mathbf{k}_\alpha;P_0)\left[ 1 + \frac{1}{L^3}\sum_{\mathbf{q}}^\Lambda
  M_L(-\mathbf{k}_\alpha,-\mathbf{q};P_0)\frac{1}{2w(\mathbf{q})}\,
  \tau_L(-\mathbf{q};P_0)\right]\, .\quad
\en
Further, using Eq.~(\ref{eq:ML-factor}) and performing Cauchy integration in Eq.~(\ref{eq:FL}), one gets:
\eq\label{eq:fin}
L^{3/2}|\langle n| J^\dagger_K(0) |0\rangle| = \biggl|\frac{g_0}{f_0}\,
\frac{1}{L^3}\sum\limits_{\mathbf{q}}^{\Lambda}\psi_L^{(n)}(-\mathbf{q})
\frac{1}{2w(\mathbf{q})}\,  \tau_L(-\mathbf{q};E_n)\biggr|\, .
\en
Now, carrying out the calculations in the infinite volume, we get:
\eq\label{eq:inf}
&&\langle \pi(\mathbf{k}_1) \pi(\mathbf{k}_2) \pi(\mathbf{k}_3);\mbox{out} |J^\dagger_K(0) |0\rangle
\nonumber\\[2mm]
&=& \frac{g_0}{f_0}\sum\limits_{\alpha=1}^{3} \tau(-\mathbf{k}_\alpha;P_0)
\left[ 1 + \int^{\Lambda}\frac{d^3\mathbf{q}}{(2\pi)^3}\phantom{.}
  M(-\mathbf{k}_\alpha,-\mathbf{q};P_0)\frac{1}{2w(\mathbf{q})}\,
  \tau(-\mathbf{q};P_0)\right]\, ,
\en
where the particle-dimer scattering amplitude $M$ is the solution of Eq.~(\ref{eq:int-Faddeev}).

Finally, comparing Eqs.~(\ref{eq:fin}) and (\ref{eq:inf}), one gets:
\eq\label{eq:ourLL}
\langle \pi(\mathbf{k}_1) \pi(\mathbf{k}_2) \pi(\mathbf{k}_3);\mbox{out} |J^\dagger_K(0) |0\rangle
=\Phi_3(\{{\bf k}\})\cdot L^{3/2}\langle n| J^\dagger_K(0) |0\rangle\, ,
\en
where the leading-order three-particle LL factor is given by:
\eq\label{eq:LL3}
\Phi_3(\{{\bf k}\})=
\pm\dfrac{\displaystyle{\sum\limits_{\alpha=1}^{3}} \tau(-\mathbf{k}_\alpha;P_0)
\left[ 1 + \displaystyle{\int^{\Lambda}}\dfrac{d^3\mathbf{q}}{(2\pi)^3}\phantom{.}
  M(-\mathbf{k}_\alpha,-\mathbf{q};P_0)\dfrac{1}{2w(\mathbf{q})}\,
  \tau(-\mathbf{q};P_0)\right]}
{\dfrac{1}{L^3}\displaystyle{\sum\limits_{\mathbf{q}}^{\Lambda}}\psi_L^{(n)}(-\mathbf{q})
  \dfrac{1}{2w(\mathbf{q})}\,  \tau_L(-\mathbf{q};E_n)}\, .
\en
The above equation implies that in the lattice measurement the box size $L$ was adjusted so that $P_0=E_n=M$ is exactly fulfilled in the rest frame of the kaon. Note also that
the numerator in Eq.~(\ref{eq:LL3}) is a complex quantity and the
Eq.~(\ref{eq:ourLL}) predicts both the real and imaginary parts of the infinite-volume matrix element, up to an overall sign. The phase of the infinite-volume decay amplitude
is determined by what can be termed the Watson theorem in the three-body case.

The equations (\ref{eq:ourLL}) and (\ref{eq:LL3}) describe
our final result. As seen, all quantities in Eq.~(\ref{eq:LL3}) can be expressed
through the couplings $C_0$ and $D_0$ which, in their turn, can be extracted from the independent
measurement of the two- and three-particle spectra. The analogy with the two-body
LL formula is now complete.

\section{Higher orders}
\label{sec:high}

For a two-particle system, the LL formula contains a single factor to all orders. This is not the
case for three particles anymore.
The situation is completely similar to the three-particle quantization condition. In this
section, we would like to briefly discuss the generalization of the approach, described
above, in the case when the 
higher-order (derivative) couplings are included in the effective Lagrangian.

We start our discussion from the two-body decays. Suppose,
the particle with a mass $M$
decays in the CM frame into two identical particles with
the mass $m$. In the infinite volume,
the physical back-to-back
momenta are then fixed by energy conservation $M=2\sqrt{m^2+k^2}$.
On the lattice, let us fix the momenta, say, along the third axis, assuming
$\mathbf{k}_1=(0,0,n)$ and $\mathbf{k}_2=(0,0,-n)$ in the units of $2\pi/L$.
Here,
$n$ is an integer number (the choice of the direction
does not matter, due to the rotational invariance). For a fixed $n$, one may adjust
$L$ so that the energy of the two-particle state
equals to the mass of the decaying particle. One then measures the finite-volume decay
matrix element exactly at this value of $L$, applies the LL formula and finally extracts the
infinite-volume matrix element one is looking for. What remains veiled in this discussion
is that one could choose different values of $n$ and $L$, so that the total energy stays
the same. In practice, this corresponds to considering the different (ground and excited) states. The matrix elements,
measured in these states, are different, and so are the pertinent LL factors. The crucial
point is that these two quantities are always correlated, so that one always extracts
the same physical infinite-volume amplitude out of the different measurements.
The mathematical reason for this correlation is that there exists only one independent
two-body decay coupling at all orders, and the finite-volume decay amplitudes in different
 states should be expressed in terms of this single coupling.

It becomes now crystal clear, what changes in case of three-particle decays. The
distribution of energies between three decay products is not fixed by the energy
conservation anymore.
This results in a non-trivial momentum dependence of the decay amplitude,
which is conveniently described by a tower of the effective couplings
$G_0,G_1,G_2,\ldots$ in the Lagrangian, multiplying the operators containing more and more spatial
derivatives. Truncating the expansion at a given order, one gets $N$ independent
couplings, which should be fixed by the measurement of $N$ linearly independent
finite-volume amplitudes. Consequently, in general, the LL factor is not a single
function. It is rather a $N\times N$ matrix, depending of the pion interaction parameters
in the two-body ($C_0,\ldots$) and three-body ($D_0,\ldots$) sectors. Using this matrix enables
one to map the results of the measurements of the matrix elements in different
states onto the couplings $G_0,G_1,G_2,\ldots$ (note that the states $n$ implicitly
depend on the momenta ${\bf k}_1,{\bf k}_2,{\bf k}_3$, which enter
the source/sink operator). At the next step, using the infinite-volume
scattering equations, it is possible to calculate pion rescattering in the final state and
express the physical decay matrix element in arbitrary kinematics. The above discussion
also shows that the extraction of the effective couplings represents a convenient
strategy in the analysis of the lattice data.

The second question, which emerges during the generalization of the approach to higher
orders, is predominantly of a technical nature. Namely, in the present formulation,
the final-state rescattering corrections in the three-particle states at higher orders are
not given in an explicitly Lorentz-invariant form. Albeit there is nothing wrong with
this in principle, an explicitly Lorentz-invariant setting in the three-particle sector
would provide a far nicer and more compact framework at higher orders, containing less effective couplings from the beginning (nothing will change at the leading order we are working in).
Note that
such a technical modification has
already been considered within an alternative formulation of the
three-body quantization condition. The modification, which boils down to the
replacement of the energy denominators by the explicitly Lorentz-invariant expressions that coincide with the former on the energy shell, has been discussed
in detail in Refs.~\cite{Briceno:2017tce,Romero-Lopez:2019qrt,Blanton:2019igq,Blanton:2020gha}. It remains to be seen, how (and whether) the similar idea can be
implemented within our approach.

\section{Conclusions}
\label{sec:concl}

\begin{itemize}

\item[i)] In the present paper, using the non-relativistic effective Lagrangian approach,
  we have derived the leading-order counterpart of the
  LL formula for three-particle decays. As in the
  two-particle case, the LL factor depends on the parameters of the pion interactions
  only (both in the two- and three-particle sectors), which can be measured
  independently from the decay matrix element in the same lattice setup.
\item[ii)]
  At higher orders, the LL factor becomes a $N\times N$ matrix, where $N$ denotes the
  number of independent couplings that describe the elementary act of the
  three-particle decay at this order. These couplings provide a convenient
  parameterization of the decay amplitude for the extraction on the lattice. The
  infinite-volume amplitudes (in an arbitrary continuum kinematics) can be calculated
  {\it a posteriori,} solving the scattering equations in the infinite volume.
\item[iii)]
  Some technical issues remain to be solved in higher orders. For example, an explicitly
  Lorentz-invariant framework would be more convenient (albeit not obligatory)
  to carry out the extraction,
  because the invariance puts stringent constraints on the possible form of the amplitude,
  reducing the number of the effective couplings needed at a given order. At the leading order,
  where the pertinent operator in the Lagrangian does not contain derivatives at all, this
  issue is not relevant. Other technical modifications concern the decays of particles with spin, partial wave
  mixing, moving frames, etc. The work in this direction is already in progress, and the
  results will be reported elsewhere.
\item[iv)] As noted already,
  the above-mentioned modifications do not affect our result, obtained at the
  leading order in the non-relativistic EFT.
  Taking into account the present state
  of lattice studies in the three-particle sector, one expects that
  in the beginning,
  all these higher-order effects will be of mainly academic interest, and the leading-order
  formula will completely suffice in the applications.

\end{itemize}

\acknowledgments

The authors would like to thank Ulf-G. Mei{\ss}ner
for interesting discussions. 
The work  was  funded in part by the Deutsche Forschungsgemeinschaft
(DFG, German Research Foundation) – Project-ID 196253076 – TRR 110.
A.~R., in addition, thanks Volkswagenstiftung 
(grant no. 93562) and the Chinese Academy of Sciences (CAS) President's
International Fellowship Initiative (PIFI) (grant no. 2021VMB0007) for the partial
financial support.

\appendix

\bibliographystyle{JHEP}

\begin{thebibliography}{99}
  
\bibitem{Lellouch:2000pv}
L.~Lellouch and M.~L\"uscher,
``Weak transition matrix elements from finite volume correlation functions,''
Commun. Math. Phys. \textbf{219} (2001) 31
[arXiv:hep-lat/0003023 [hep-lat]].


\bibitem{Kim:2005gf}
C.~h.~Kim, C.~T.~Sachrajda and S.~R.~Sharpe,
``Finite-volume effects for two-hadron states in moving frames,''
Nucl. Phys. B \textbf{727} (2005) 218
[arXiv:hep-lat/0507006 [hep-lat]].

\bibitem{Christ:2005gi}
N.~H.~Christ, C.~Kim and T.~Yamazaki,
``Finite volume corrections to the two-particle decay of states with non-zero momentum,''
Phys. Rev. D \textbf{72} (2005) 114506
[arXiv:hep-lat/0507009 [hep-lat]].

\bibitem{Hansen:2012tf}
M.~T.~Hansen and S.~R.~Sharpe,
``Multiple-channel generalization of Lellouch-L\"uscher formula,''
Phys. Rev. D \textbf{86} (2012) 016007
[arXiv:1204.0826 [hep-lat]].

\bibitem{Bernard:2012bi}
V.~Bernard, D.~Hoja, U.-G.~Mei{\ss}ner and A.~Rusetsky,
``Matrix elements of unstable states,''
JHEP \textbf{09} (2012) 023
[arXiv:1205.4642 [hep-lat]].


\bibitem{Abbott:2020hxn}
R.~Abbott \textit{et al.} [RBC and UKQCD],
``Direct CP violation and the $\Delta I=1/2$ rule in $K\to\pi\pi$ decay from the standard model,''
Phys. Rev. D \textbf{102} (2020) 054509
[arXiv:2004.09440 [hep-lat]].

\bibitem{Briceno:2015csa}
R.~A.~Brice\~no and M.~T.~Hansen,
``Multichannel 0 $\to$ 2 and 1 $\to$ 2 transition amplitudes for arbitrary spin particles in a finite volume,''
Phys. Rev. D \textbf{92} (2015) 074509
[arXiv:1502.04314 [hep-lat]].https://inspirehep.net/literature/1302412

\bibitem{Briceno:2014uqa}
R.~A.~Brice\~no, M.~T.~Hansen and A.~Walker-Loud,
``Multichannel 1 $\rightarrow$ 2 transition amplitudes in a finite volume,''
Phys. Rev. D \textbf{91} (2015)  034501
[arXiv:1406.5965 [hep-lat]].

\bibitem{Meyer:2011um}
H.~B.~Meyer,
``Lattice QCD and the Timelike Pion Form Factor,''
Phys. Rev. Lett. \textbf{107} (2011) 072002
[arXiv:1105.1892 [hep-lat]].






\bibitem{Hansen:2014eka}
  M.~T.~Hansen and S.~R.~Sharpe,
  ``Relativistic, model-independent, three-particle quantization condition,''
  Phys.\ Rev.\ D {\bf 90} (2014) 116003 [arXiv:1408.5933 [hep-lat]].

\bibitem{Hansen:2015zga}
  M.~T.~Hansen and S.~R.~Sharpe,
  ``Expressing the three-particle finite-volume spectrum in terms of the three-to-three scattering amplitude,''
  Phys.\ Rev.\ D {\bf 92} (2015) 114509 [arXiv:1504.04248 [hep-lat]].



\bibitem{Hammer:2017uqm}
H.~W.~Hammer, J.~Y.~Pang and A.~Rusetsky,
``Three-particle quantization condition in a finite volume: 1. The role of the three-particle force,''
JHEP \textbf{09} (2017) 109
[arXiv:1706.07700 [hep-lat]].


\bibitem{Hammer:2017kms}
H.~W.~Hammer, J.~Y.~Pang and A.~Rusetsky,
``Three particle quantization condition in a finite volume: 2. general formalism and the analysis of data,''
JHEP \textbf{10} (2017) 115
[arXiv:1707.02176 [hep-lat]].

  
\bibitem{Mai:2017bge}
  M.~Mai and M.~D\"oring,
  ``Three-body Unitarity in the Finite Volume,''
  Eur.\ Phys.\ J.\ A {\bf 53} (2017)  240
  [arXiv:1709.08222 [hep-lat]].


\bibitem{Beane:2007es}
S.~R.~Beane, W.~Detmold, T.~C.~Luu, K.~Orginos, M.~J.~Savage and A.~Torok,
``Multi-Pion Systems in Lattice QCD and the Three-Pion Interaction,''
Phys. Rev. Lett. \textbf{100} (2008) 082004
[arXiv:0710.1827 [hep-lat]].

\bibitem{Horz:2019rrn}
B.~H\"orz and A.~Hanlon,
``Two- and three-pion finite-volume spectra at maximal isospin from lattice QCD,''
Phys. Rev. Lett. \textbf{123} (2019) 142002
[arXiv:1905.04277 [hep-lat]].

\bibitem{Culver:2019vvu}
  C.~Culver, M.~Mai, R.~Brett, A.~Alexandru and M.~D\"oring,
 ``Three pion spectrum in the $I=3$ channel from lattice QCD,''
Phys. Rev. D \textbf{101} (2020) 114507
[arXiv:1911.09047 [hep-lat]].

\bibitem{Fischer:2020jzp}
  M.~Fischer, B.~Kostrzewa, L.~Liu, F.~Romero-L\'opez, M.~Ueding and C.~Urbach,
  ``Scattering of two and three physical pions at maximal isospin from lattice QCD,''
[arXiv:2008.03035 [hep-lat]].

\bibitem{Blanton:2019vdk}
T.~D.~Blanton, F.~Romero-L\'opez and S.~R.~Sharpe,
``$I=3$ Three-Pion Scattering Amplitude from Lattice QCD,''
Phys. Rev. Lett. \textbf{124} (2020) 032001
[arXiv:1909.02973 [hep-lat]].

\bibitem{Hansen:2020otl}
M.~T.~Hansen, R.~A.~Brice\~no, R.~G.~Edwards, C.~E.~Thomas and D.~J.~Wilson,
``The energy-dependent $\pi^+ \pi^+ \pi^+$ scattering amplitude from QCD,''
[arXiv:2009.04931 [hep-lat]].

\bibitem{Alexandru:2020xqf}
A.~Alexandru, R.~Brett, C.~Culver, M.~D\"oring, D.~Guo, F.~X.~Lee and M.~Mai,
``Finite-volume energy spectrum of the $K^-K^-K^-$ system,''
[arXiv:2009.12358 [hep-lat]].


\bibitem{Romero-Lopez:2018rcb}
F.~Romero-L\'opez, A.~Rusetsky and C.~Urbach,
``Two- and three-body interactions in $\varphi ^4$ theory from lattice simulations,''
Eur. Phys. J. C \textbf{78} (2018) 846
[arXiv:1806.02367 [hep-lat]].

\bibitem{Romero-Lopez:2020rdq}
F.~Romero-L\'opez, A.~Rusetsky, N.~Schlage and C.~Urbach,
``Relativistic $N$-particle energy shift in finite volume,''
[arXiv:2010.11715 [hep-lat]].

\bibitem{Hansen:2020zhy}
M.~T.~Hansen, F.~Romero-L\'opez and S.~R.~Sharpe,
  ``Generalizing the relativistic quantization condition to include all three-pion isospin channels,''
JHEP \textbf{07} (2020) 047
[arXiv:2003.10974 [hep-lat]].

\bibitem{Muller:2020vtt}
F.~M\"uller, A.~Rusetsky and T.~Yu,
``Finite-volume energy shift of the three-pion ground state,''
[arXiv:2011.14178 [hep-lat]].



\bibitem{Hansen:2021ofl}
M.~T.~Hansen, F.~Romero-L\'opez and S.~R.~Sharpe,
``Decay amplitudes to three hadrons from finite-volume matrix elements,''
[arXiv:2101.10246 [hep-lat]].


\bibitem{Agadjanov:2014kha}
A.~Agadjanov, V.~Bernard, U.-G.~Mei{\ss}ner and A.~Rusetsky,
``A framework for the calculation of the $\Delta N \gamma^*$
transition form factors on the lattice,''
Nucl. Phys. B \textbf{886} (2014), 1199
[arXiv:1405.3476 [hep-lat]].

\bibitem{Agadjanov:2016fbd}
A.~Agadjanov, V.~Bernard, U.-G.~Mei{\ss}ner and A.~Rusetsky,
``The $B\to K^*$ form factors on the lattice,''
Nucl. Phys. B \textbf{910} (2016) 387
[arXiv:1605.03386 [hep-lat]].

\bibitem{Briceno:2015dca}
R.~A.~Briceno, J.~J.~Dudek, R.~G.~Edwards, C.~J.~Shultz, C.~E.~Thomas and D.~J.~Wilson,
``The resonant $\pi^+\gamma\to\pi^+\pi^0$ amplitude from Quantum Chromodynamics,''
Phys. Rev. Lett. \textbf{115} (2015) 242001
[arXiv:1507.06622 [hep-ph]].


\bibitem{Briceno:2016kkp}
R.~A.~Brice\~no, J.~J.~Dudek, R.~G.~Edwards, C.~J.~Shultz, C.~E.~Thomas and D.~J.~Wilson,
``The $\pi\pi\to\pi\gamma^\star$ amplitude and the resonant $\rho\to\pi\gamma^\star$ transition from lattice QCD,''
Phys. Rev. D \textbf{93} (2016) 114508
[arXiv:1604.03530 [hep-ph]].





\bibitem{Colangelo:2006va}
G.~Colangelo, J.~Gasser, B.~Kubis and A.~Rusetsky,
``Cusps in $K\to 3 \pi$ decays,''
Phys. Lett. B \textbf{638} (2006) 187
[arXiv:hep-ph/0604084 [hep-ph]].


\bibitem{Bissegger:2008ff}
M.~Bissegger, A.~Fuhrer, J.~Gasser, B.~Kubis and A.~Rusetsky,
``Radiative corrections in $K\to 3\pi$ decays,''
Nucl. Phys. B \textbf{806} (2009) 178
[arXiv:0807.0515 [hep-ph]].

\bibitem{Bissegger:2007yq}
M.~Bissegger, A.~Fuhrer, J.~Gasser, B.~Kubis and A.~Rusetsky,
``Cusps in $K_L\to 3\pi$ decays,''
Phys. Lett. B \textbf{659} (2008) 576
[arXiv:0710.4456 [hep-ph]].

\bibitem{Gullstrom:2008sy}
C.~O.~Gullstr\"om, A.~Kup\'s\'c and A.~Rusetsky,
``Predictions for the cusp in $\eta\to 3\pi^0$ decay,''
Phys. Rev. C \textbf{79} (2009) 028201
[arXiv:0812.2371 [hep-ph]].

\bibitem{Schneider:2010hs}
S.~P.~Schneider, B.~Kubis and C.~Ditsche,
``Rescattering effects in $\eta\to 3\pi$ decays,''
JHEP \textbf{02} (2011) 028
[arXiv:1010.3946 [hep-ph]].

\bibitem{Kubis:2009sb}
B.~Kubis and S.~P.~Schneider,
``The Cusp effect in $\eta'\to\eta\pi\pi$ decays,''
Eur. Phys. J. C \textbf{62} (2009) 511
[arXiv:0904.1320 [hep-ph]].


\bibitem{Gasser:2011ju}
J.~Gasser, B.~Kubis and A.~Rusetsky,
``Cusps in $K\to 3\pi$ decays: a theoretical framework,''
Nucl. Phys. B \textbf{850} (2011) 96
[arXiv:1103.4273 [hep-ph]].



\bibitem{Kaplan:1996nv}
D.~B.~Kaplan,
``More effective field theory for nonrelativistic scattering,''
Nucl. Phys. B \textbf{494} (1997) 471
[arXiv:nucl-th/9610052 [nucl-th]].

\bibitem{Bedaque:1998kg}
P.~F.~Bedaque, H.~W.~Hammer and U.~van Kolck,
``Renormalization of the three-body system with short range interactions,''
Phys. Rev. Lett. \textbf{82} (1999) 463
[arXiv:nucl-th/9809025 [nucl-th]].

\bibitem{Bedaque:1998km}
P.~F.~Bedaque, H.~W.~Hammer and U.~van Kolck,
``The Three boson system with short range interactions,''
Nucl. Phys. A \textbf{646} (1999) 444
[arXiv:nucl-th/9811046 [nucl-th]].

\bibitem{Doring:2018xxx}
M.~D\"oring, H.~W.~Hammer, M.~Mai, J.~Y.~Pang, A.~Rusetsky and J.~Wu,
``Three-body spectrum in a finite volume: the role of cubic symmetry,''
Phys. Rev. D \textbf{97} (2018) 114508
[arXiv:1802.03362 [hep-lat]].



\bibitem{Kreuzer:2010ti}
  S.~Kreuzer and H.-W.~Hammer,
  ``The Triton in a finite volume,''
  Phys.\ Lett.\ B {\bf 694} (2011) 424 [arXiv:1008.4499 [hep-lat]].

\bibitem{Kreuzer:2009jp}
 S.~Kreuzer and H.-W.~Hammer,
  ``On the modification of the Efimov spectrum in a finite cubic box,''
  Eur.\ Phys.\ J.\ A {\bf 43} (2010) 229 [arXiv:0910.2191 [nucl-th]].

\bibitem{Kreuzer:2008bi}
  S.~Kreuzer and H.-W.~Hammer,
  ``Efimov physics in a finite volume,''
  Phys.\ Lett.\ B {\bf 673} (2009) 260 [arXiv:0811.0159 [nucl-th]].

\bibitem{Kreuzer:2012sr}
  S.~Kreuzer and H.-W.~Grie{\ss}hammer,
  ``Three particles in a finite volume: The breakdown of spherical symmetry,''
  Eur.\ Phys.\ J.\ A {\bf 48} (2012) 93 [arXiv:1205.0277 [nucl-th]].


\bibitem{Polejaeva:2012ut}
K.~Polejaeva and A.~Rusetsky,
``Three particles in a finite volume,''
Eur. Phys. J. A \textbf{48} (2012) 67
[arXiv:1203.1241 [hep-lat]].


  
\bibitem{Briceno:2017tce}
R.~A.~Brice\~no, M.~T.~Hansen and S.~R.~Sharpe,
``Relating the finite-volume spectrum and the two-and-three-particle $S$ matrix for relativistic systems of identical scalar particles,''
Phys. Rev. D \textbf{95} (2017) 074510
[arXiv:1701.07465 [hep-lat]].

\bibitem{Romero-Lopez:2019qrt}
F.~Romero-L\'opez, S.~R.~Sharpe, T.~D.~Blanton, R.~A.~Brice\~no and M.~T.~Hansen,
``Numerical exploration of three relativistic particles in a finite volume including two-particle resonances and bound states,''
JHEP \textbf{10} (2019) 007
[arXiv:1908.02411 [hep-lat]].

\bibitem{Blanton:2019igq}
T.~D.~Blanton, F.~Romero-L\'opez and S.~R.~Sharpe,
``Implementing the three-particle quantization condition including higher partial waves,''
JHEP \textbf{03} (2019) 106
[arXiv:1901.07095 [hep-lat]].

\bibitem{Blanton:2020gha}
T.~D.~Blanton and S.~R.~Sharpe,
``Alternative derivation of the relativistic three-particle quantization condition,''
Phys. Rev. D \textbf{102} (2020) 054520
[arXiv:2007.16188 [hep-lat]].
  


\end{thebibliography}

\end{document}